\documentstyle[epsfig]{article}

\if twoside
\else
\fi

\begin{document}
\pagestyle{plain}
\sloppy
\newcommand{\ba}{\begin{eqnarray}}
\newcommand{\ea}{\end{eqnarray}}
\newcommand{\bna}{\begin{array*}}
\newcommand{\ena}{\end{array*}}
\newcommand{\be}{\begin{equation}}
\newcommand{\ee}{\end{equation}}
\newcommand{\bZ}{{\bf Z}}
\newcommand{\ab}{\\[1mm]}
\def\m{\mu}
\def\n{\nu}
\def\o{\omega}
\def\p{\partial}
\def\ph{\phi}
\def\ps{\psi}
\def\a{\alpha}
\def\c{\chi}
\def\d{\delta}
\def\ti{\tilde}
\def\t{\tau}
\def\e{\epsilon}
\def\g{\gamma}
\def\b{\beta}
\def\k{\kappa}
\def\l{\lambda}
\def\s{\sigma}
\baselineskip 12pt
\large {\bf \noindent The structure of the QED-Vacuum and Electron - Positron Pair Production in Super - Intense, pulsed Laser Fields}
\vspace*{2cm}\\
\hspace*{3cm}
K.~Dietz and M.~Pr"obsting \\
\vspace*{1cm}\\ 
\hspace*{2cm} \small{{\em Physikal.~Institut der Universit"at Bonn, Nu"sallee 12,\\
\hspace*{2cm} 53115 Bonn, Germany} \\ \\ {\em\hspace*{2cm}
e-mail: dietz@pib1.physik.uni-bonn.de}} \\ \\ \hspace*{5cm}
\vspace*{1cm} {\oddsidemargin 1.35cm}
\begin{abstract}
\renewcommand{\baselinestretch}{2.0}
\small\normalsize
We discuss electron - positron pair - production by super - intense,
short laser pulses off the physical vacuum state locally deformed
by (stripped) nuclei with large nuclear charges. Consequences of
non - perturbative vacuum polarisation resulting from such a deformation
are shortly broached. Production probabilities per pulse are calculated.
\end{abstract}
\newpage
\renewcommand{\baselinestretch}{2.0}
\small\normalsize
Vacuum fluctuations entail observable effects : local deformation
of the translationally invariant vacuum state leads to vacuum
polarisation which is amenable to experimentation. The classic
example for such a situation is the Casimir effect where two parallel
plates deform the vacuum state and thus give rise to an attractive
potential. Much more efficient polarisations of the vacuum are produced
by point - like charges, e.g. stripped nuclei with large Z.\\
One of the effects produced by such local Coulomb deformations can be
observed in \\
heavy - ion collisions : non - perturbative vacuum
polarisation prohibits the intrusion of strongly bound electron
energy levels into the Dirac sea spectrum and thus precludes positron
production via the 'charged vacuum' mechanism proposed by Greiner et al. \cite{size}.
The recently observed absense \cite{experiment} of such positron emission processes is
in accord with this mechanism predicting repulsive forces which prevent electron
levels from penetrating into the Dirac sea even beyond 'critical'
nuclear Coulomb charges.\\
In this letter we shall address yet another physical consequence of
local vacuum \\
deformation : multi - photon production of electron -
positron pairs in super - intense laser fields. As we shall see
multi - photon transitions involving $10^{5}-10^{6}$ photons occur
with measurable probability even at laser intensities near the reach
of experimentation. Super - intense, pulsed laser beams, hence, might
be used as probes in an investigation of the structure of the
physical vacuum which in itself is an interesting aspect of such pair
production processes.\\
Conceptually, the calculation of pair - production probabilities follows
clear guide - lines. Let us start with a short description of
non - perturbative vacuum polarisation. To encode the many - body
nature of the problem we use a quantum field theoretic approach and
write
\be
H=\int d^3 x \varphi ^* (x,t)\left( -i\vec{\alpha}\cdot\vec{\partial}+\beta
m+V_{ext}\right) \varphi (x,t)
\ee
where $\varphi (x,t)$ denotes the electron Fermi - field and the
notation for the Dirac operator is chosen as in \cite{Standard}; for
the moment
\be
V_{ext}=-\frac{Z\,e^2}{|\vec{x}|}
\ee
is the potential, seen by the electron, produced by a point - like nucleus
of charge $Z\,e$ (in numerical calculations finite size effects are taken
into account as e.g. in \cite{size} ). We expand
\be\label{expansion}
\varphi (x,t)=\sum_{\alpha} \Psi_{\alpha}(x)\,e^{-i\epsilon_{\alpha}t}\,
a_{\alpha},
\ee
where
\be\label{eigen}
H_{D}\Psi_{\alpha}\equiv \left( -i\vec{\alpha}\vec{\partial}+\beta
m+V_{ext}\right) \Psi_{\alpha} = \epsilon_{\alpha}\Psi_{\alpha},
\ee
and obtain for the ( symmetrized ) vacuum energy
\be
E_{vac}=-\frac{1}{2}\left(\sum_{\epsilon_{\alpha}>-m} \epsilon_{\alpha}
-\sum_{\epsilon_{\alpha}<-m} \epsilon_{\alpha}\right),
\ee
i.e. the energy of the state ( $N$ is a normalisation factor )
\be
\Psi_{D}=N\prod_{\epsilon_{\alpha}<-m} a_{\alpha}^{\dagger}\Psi_{0},
\ee
$\Psi_{0}$ is the Fock vacuum. The polarisation charge density
is obtained as
\be
\rho_{vac}(\vec{x})=e\frac{\delta E_{vac}}{\delta V_{ext}(\vec{x})}=
-\frac{1}{2}\left(\sum_{\epsilon_{\alpha}>-m} |\Psi_{\alpha}(\vec{x})|^2
-\sum_{\epsilon_{\alpha}<-m} |\Psi_{\alpha}(\vec{x})|^2 \right).
\ee
Clearly,
\be
\rho_{vac}=0
\ee
for $V_{ext}=0$. Upon switching on the Coulomb potential ( screened at
large distances ), we find
\be
\rho_{vac}\ne 0 ;
\ee
the total charge, however, is a deformation invariant \cite{A1} and stays
zero modulo integer multiples of $e$. As we pointed out \cite{A2}, the
total charge
\be
\eta_{vac}=\int d^3x \rho_{vac}(\vec{x})
\ee
jumps by $-e$ at the positron threshold $-m$ and thus yields a repulsive
potential preventing the 'about to intrude' bound electron level from
producing a 'charged vacuum' state by entering the Dirac sea continuum (DSC)\footnote{The real number $\Delta$ introduced in \cite{Standard} and
\cite{A1} for mathematical consistency has been set $\Delta =0$ (as in
all our previous numerical calculations). Of course, this expresses the
original definition of antiparticles formulated by Dirac.}.\\
Our next step is now to study the effect of a twofold source of vacuum\\
deformation : in addition to the large Coulomb field originating from
a nuclear charge we employ a super-intense pulsed laser field to induce
further vacuum deformation.
Introducing now a linearly, in z - direction, polarised, pulsed laser field
\be
\vec{A}=\int^t dt'\,\,\lambda (t')\cos (\omega t')\vec{e}_{z},
\ee
with the pulsed $(\mbox{intensity})^{\frac{1}{2}}$ $\lambda (t)$ 
and frequency $\omega$, we have, employing the dipole approximation
(which can be seen to be valid, comparing the appropriate scales with
the laser wavelengths under consideration),
\be\label{Vext}
V_{ext}(\vec{x},t)=-\frac{Z\,e^2}{|\vec{x}|}+e\,\lambda (t)\vec{x}\cdot
\vec{e}_{z}\cos (\omega t).
\ee
The dynamics of pulsed laser interactions is efficiently described in a
picture of dressed states which now build up a dressed Dirac vacuum. Pulsing the
laser beam now induces transitions which we propose to observe as
$e^+$-$e^-$ pairs. Dressed states can be represented as Floquet states
\begin{eqnarray}
\left(H_{D}-i\partial_{t}\right)\tilde{\Psi}_{\alpha}(x,t) & = & \tilde{\epsilon}
_{\alpha}\tilde{\Psi}_{\alpha}(x,t) \nonumber \\
\tilde{\Psi}_{\alpha}(x,t) & = & \tilde{\Psi}_{\alpha}(x,t+\frac{2\pi}{\omega})
\end{eqnarray}
which are used to decompose the electron field as in (\ref{expansion})
\be
\varphi (x,t)=\sum_{\alpha} \tilde{\Psi}_{\alpha}(x,t)e^{-i\tilde{\epsilon}_{\alpha}t}
\tilde{a}_{\alpha}
\ee
thus defining creation and annihilation operators for dressed states.\\
The dressed Dirac vacuum state is then given as \footnote{The states
$a_{\alpha}^{\dagger}\Psi_{0}$ are for $\lambda \rightarrow 0$
continuously connected to the eigenstates defined in (\ref{eigen})}
\be
\Psi_{D}^{dressed}=N\prod_{\tilde{\epsilon}_{\alpha}<-m}\tilde{a}_{\alpha}
^{\dagger}\Psi_{0}.
\ee
More precisely speaking, the spectrum $\{\tilde{\epsilon}_{\alpha}\}$
is continuous and covers the real axis, bound electron states turn into broad
resonances whose widths increases with the laser intensity.\\
In the dressed state picture, a reaction with a pulsed laser beam is described
as a transport process in which the state $\{\epsilon_{\alpha},\Psi_{\alpha}\}$
is dressed and transported \cite{A3} along the pulse envelope $\lambda (t)$;
during this transport transitions to dressed resonances occur which in
the switch - off phase of the pulse are turned into occupied bound or
scattering electron states : after the pulse we observe a bound or scattered
electron and a hole in DSC : a positron - electron pair has been produced.
Of course, double pair production, leaving two holes and either bound or
scattered electrons, takes place, the corresponding states will have
non - vanishing amplitudes in the pulse - transported state with
notably suppressed probabilities, of course.\\
Two circumstances enhance the probability for pair production to take
place and the possibility of experimental detections : choosing large
$Z$ diminishes the distance of DSC to the lowest electron bound state and, hence,
favours the multi - photon pair production process; the resonance nature of
dressed bound electron states entails during the reaction
a non - negligible energy band width of produced pairs and, thus, an
energy - dispersed pair - production spectrum.\\
There are methods \cite{A4} which aim at the calculation of resonance
width and position which cannot be expected to lead to reliable results
for laser intensities which we envisage. Such a calculation would be
desirable for a detailed, qualitative discussion of the dynamics of
pair-production
processes; we however have to forgo such an analysis. Nonetheless, we
include all these effects implicitly by using the equation of motion
to follow the time-development of the state vector of the system from
the beginning to the end of the pulse.\\
We calculate the probability per pulse $P_{p}$ for transitions
\be\label{reaction}
|\alpha>_{E=-\epsilon_{\alpha}>m}\stackrel{\rm laser~pulse}{\longrightarrow}
|\beta>_{\epsilon_{\beta}>-m}
\ee
by numerically solving the initial state problem
\be\label{inistat}
\Psi (x,t)_{|_{t=0}}=|\vec{p}>_{E=-\epsilon_{\alpha}>m}
\ee
for the Dirac equation
\be\label{Diracequa}
\left(H_{D}-i\partial_{t}\right)\tilde{\Psi}(x,t)=0
\ee
where $|\alpha>_{E=-\epsilon_{\alpha}>m}$ is the initial positron scattering
state (suitably redefining negative energy state quantum numbers \cite{A2})
at $t=0$ just before the switch-on of the laser pulse.
The state vector $\tilde{\Psi}(x,t)$, taken for times $t$ larger than the duration
of the laser pulse, is assumed to have, following the discussion given above, only single - electron state components with amplitudes
$a_{\alpha}^{\beta}(t_{f}),\epsilon_{\alpha}>-m$ besides the initial
state (\ref{inistat}). That is to say that the effective space in
which our calculations are performed is characterised by the projection
operator
\be
P_{eff}=|\alpha><\alpha|_{|_{E=-\epsilon_{\alpha}}}+\sum |\beta ><\beta|_{|_{\epsilon_{\beta}>-m}},
\ee
where $|\alpha>$ is the already mentioned positron - scattering state
and $\{|\beta >\}$ are \\
single - electron bound and scattering states.
A moment's thought shows that this selection of
states is dictated by the Pauli principle incorporated in the field
theoretic picture shortly sketched above. Needless to say, by solving
in this way the initial value problem in a time interval containing
switch-on and switch-off of the laser beam one fully describes the dynamics
(restricted to one-electron (positron) states); in particular the finite
width effects discussed above are accounted for.\\
Choosing $t_{f}$ as a time shortly after the pulse we finally calculate
the probability density
\be\label{Probdens}
P_{p}=\frac{\sum_{\alpha,\epsilon_{\alpha}=-E}\sum_{\nu,\epsilon_{\nu}>-m}
|a_{\alpha}^{\nu}(t_{f})|^2}{\sum_{\alpha,\epsilon_{\alpha}=-E}\,1},
\ee
including the average over (positron) states with energy $E$. The number
$n(E)$ of positrons produced in reaction (\ref{reaction}) to be
observed in the interval $[E,E+dE]$ is given as
\be
n(E)=P_{p}(E)\rho(E)dE
\ee
where $\rho(E)$ is the density of states in the DSC.\\
To give a quantitative idea of the possibility of detecting electron - positron
pairs in experiments we present a calculation \footnote{Numerical calculations
were done using the methods outlined in \cite{A1} : we used a basis set
of maximally 1600 ($j=\frac{1}{2},...,\frac{7}{2}$) eigenfunctions of
the Coulomb Dirac problem to write the time dependent Dirac equation
as set of ordinary differential equations which, in turn, were solved
by the predictor-corrector routine by Shampine and Gordon \cite{SG}.
Careful checks were employed to ascertain that the dimension of our
reference space was sufficiently large (actually it was seen that stable
results already appeared at a significantly lower dimension)
to guarantee numbers of the required accuracy.} for a conservative case :
we take an ultra short pulse with 20 cycles and a photon energy of 5eV
and calculate $P_{p}(E)$ for nuclei with $Z=70$ and $Z=100$. The results
are shown in Fig.1. Realistic pulses encompass several thousand
cycles and the $P_{p}(E)$'s shown should be increased by at least an order of
magnitude - we conclude that cross sections, i.e. probability per pulse,
are within the detection capabilities of experimental setups for
sufficiently dense samples of stripped high $Z$ nuclei - it is beyond
the judgement of the authors to estimate the feasibility of
preparing such samples.\\
The measurement of pair production rates is highly desirable since
interesting insight into the structure of the physical vacuum could be
provided in this way.

\newpage
\section*{Figure Caption}
{\bf Fig. 1 :} The probability density $P_{p}(E)$ ( see (\ref{Probdens}))
as a function of $\lambda_{max}$, $\lambda_{max}$
being the maximal field strength of the pulse $\lambda (t)$,
\be
\lambda (t)=\lambda_{max}\sin ^{2}(\frac{\pi t}{T_{p}}).
\ee
The parameters in the numerical calculation are
\begin{itemize}
\item $T_{p}=20$ (cycles of laser period)
\item $\omega =5$ eV
\item $Z=70$ and $Z=100$
\item $m=511$ keV
\item $E=513$ keV
\end{itemize}
\newpage

\oddsidemargin 0.2cm

\hskip -1.5cm
\vskip 6cm
\leftline{{\Large $P_{p}$}}
\vskip -3cm
\vspace*{-10cm}
\begin{figure}[h]
    \vbox{
      \hbox{
        \psfig{figure=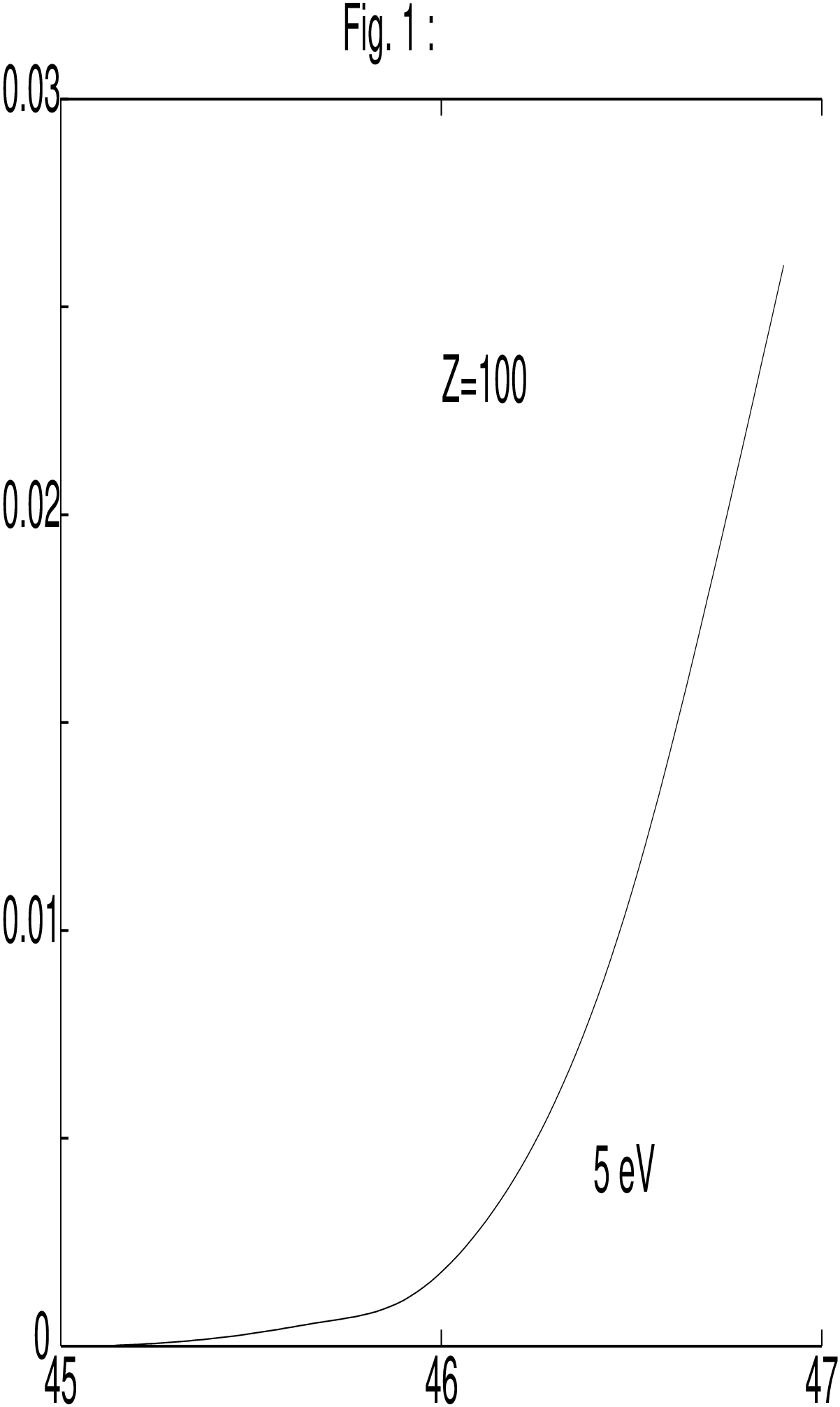,width=.48\textwidth}
        \psfig{figure=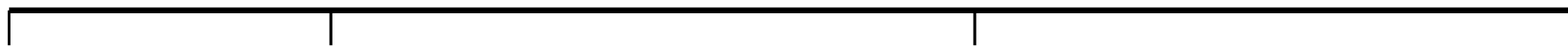,width=.48\textwidth}
        }
      }
\end{figure}
\vskip 13.0cm
\centerline{{\Large $\lambda_{max}\,\,[10^{3}a.u.]$}}
\end{document}